\documentclass{svproc}
\usepackage{url}
\usepackage{graphicx}
\usepackage{comment}
\usepackage{float}
\usepackage{multirow}

\begin{document}
\mainmatter              
\title{Data Sharing at the Edge of the Network: A Disturbance Resilient Multi-modal ITS }
%
\titlerunning{Data Sharing at the Edge of the Network}  
%
\author{Igor Mikolasek\inst{1} \and Saeedeh Ghanadbashi\inst{2} \and Nima Afraz\inst{2} \and Fatemeh Golpayegani\inst{2}
}
\authorrunning{Igor Mikolasek et al.} 
%
\tocauthor{Igor Mikolasek, Saeedeh Ghanadbashi, Nima Afraz,  Fatemeh Golpayegani}
\institute{Transport Research Centre CDV, Brno, Czech Republic,\\
\email{igor.mikolasek@cdv.cz},
\and
School of Computer Science, University College Dublin, Ireland}

\maketitle              

\begin{abstract}

Mobility-as-a-Service (MaaS) is a paradigm that  encourages the shift from private cars to more sustainable alternative mobility services. MaaS provides services that enhances  and enables multiple modes of transport to operate seamlessly and bringing Multimodal Intelligent Transport Systems (M-ITS) closer to reality. This requires sharing and integration of data collected from multiple sources including modes of transports, sensors, and  end-users' devices to allow a seamless and integrated services especially during unprecedented  disturbances. 
This paper discusses the interactions among transportation modes, networks, potential disturbance scenarios, and adaptation strategies to mitigate their impact on MaaS. We particularly  discuss the need to share data between the modes of transport and relevant entities that are at the vicinity of each other, taking advantage of edge computing technology to avoid any latency due to communication to the cloud and privacy concerns. However, when sharing at the edge, bandwidth, storage, and computational limitations must be considered.
\end{abstract}

\section{Introduction}
The concept of Mobility-as-a-Service (MaaS) is gaining interest due to increased awareness of sustainable mobility and the adoption of Intelligent Transportation Systems (ITS) \cite{esztergar2020exploring,golpayegani2022intelligent}. MaaS aims to shift people from cars to more sustainable modes of transportation, relying on ITS for efficient infrastructure use, real-time data provision, and user-guided routing. Presently, public transport (PT) systems commonly use ITS for real-time data on arrival times and vehicle locations. Similarly, bike/scooter-sharing services provide device location data, and traffic signals can optimize intersection throughput based on traffic conditions. These ITS systems often lack cooperation among different transportation modes, limiting their effectiveness particularly at the times of disturbances. Multimodal ITS (M-ITS) \cite{qureshi2013survey} can enhance MaaS, where different modes can support or substitute  one another \cite{mnif2018immune,yang2019real,jevinger2019potentials}. This paper provides an overview of transportation modes, networks, and their interactions, focusing on disturbances and how M-ITS can mitigate them. M-ITS concept can incorporate edge computing paradigm \cite{safavifar2021adaptive} to place the processing of data closer to its source. This will address privacy concerns, reduce data processing delays, enable decentralize computation, and eliminate costly centralized systems. Using edge computing paradigm,  decisions are made in a  decentralized manner, relying on data from the local transport network, onboard vehicle sensors, and information from nearby vehicles and infrastructure. Identifying data-sharing needs, and the data flow among systems and vehicles during disturbances is essential. This paper outlines such possible disturbances and their impact on different modes and suggests coordinated impact-mitigation strategies for the M-ITS-based MaaS system.

\section{Transportation Modes and Networks}

\begin{figure}[htbp]
\centerline{
\includegraphics[width=1\linewidth]{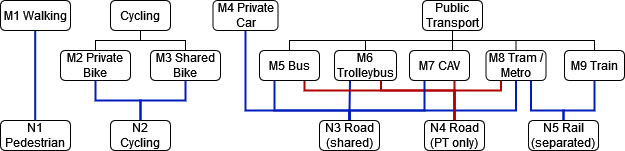}}
\caption{Overview of fundamental transportation modes and networks employed. The indices $M_i$ and $N_i$ are employed for mode and network respectively and specifying the pairings between the two in Fig. \ref{fig:effects of disturbances on different combinations}.}
\label{fig:transport modes and networks}
\end{figure}

To facilitate data sharing between multiple modes of transport, we first need to understand their dependencies and how they interact with one another. This paper focuses on transport in urban areas, although most modes can operate beyond. Expanding M-ITS and MaaS to intercity or interstate levels would involve considering additional modes, especially air transport that are out of scope for this paper. To do so, first, we model the dependencies  between transportation modes and their infrastructure networks in urban areas.\footnote{ The commonly available transportation modes provided is not exhaustive – each category can represent multiple modes of similar nature.}
Transport modes can operate on various networks, some shared and others exclusive. Networks often share infrastructure, for example roads are shared between cars, buses, and cyclists. They can be modeled collectively, with parameters defining mode usage. Multimodal modes indicate possible mode transitions. Figure \ref{fig:transport modes and networks} presents an overview of the modes noted as $M_i$ and the networks they use, noted as $N_i$.

\vspace{-1em}
\vspace{-2em}

\vspace{-10pt}
\section{Service Disturbances to Mode-Network Combinations and M-ITS Adaptation Strategies}
\begin{figure}[htbp]
\centerline{
\includegraphics[width=1\linewidth]{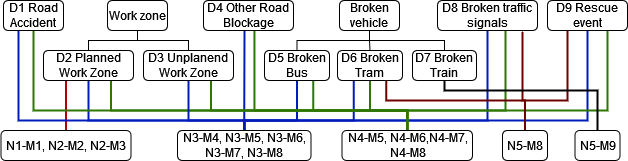}}
\caption{Effects of different sources of disturbance on different combinations of modes and networks. }
\label{fig:effects of disturbances on different combinations}
\end{figure}

Unplanned events can disrupt normal traffic operations, impacting system performance and requiring adaptations to limit such impacts.  Data sharing and collective intelligence can inform about and address the challenges related to such disturbances . Five main causes of disturbances are identified, impacting various mode-network combinations (see Fig. \ref{fig:effects of disturbances on different combinations}). Here, only cases of possible direct impacts (need to re-route) are considered. Secondary impacts due to possible relocation of transport demand are considered later for individual disturbance scenarios. In Table \ref{tab:list of detection options}, we present a comprehensive overview of individual disturbances, providing further insights into potential data sources for detection, strategies for adaptation, and special cases that might necessitate specific responses.

\vspace{-1em}
\vspace{-1em}

Below we provide some examples of disturbances and scenarios to handle them. For example, ``Planned Work Zone'' along the pedestrian network (N1-M1) might result in disturbance for  users with wheelchairs. Such work zones are not commonly recorded in a machine-readable format. This is especially true if they are unplanned, hence this case is omitted in the schema in Fig \ref{fig:effects of disturbances on different combinations}. 
Pedestrians and cyclists have the advantage of being very agile and can bypass most obstacles on the network with acceptable delay and they are not connected to any ITS service, except shared bikes. Therefore, except for planned work zones, they are not considered primary stakeholders. These modes are considered here mostly as transition modes and as an adaptation strategy in case of disturbance.
\begin{table}[]
\footnotesize
\centering
\caption{List of detection options, adaptation strategies, and specific cases for the basic identified disturbances.}
\label{tab:list of detection options}
\begin{tabular}{|p{0.01\linewidth}|p{0.34\linewidth}|p{0.347\linewidth}|p{0.18\linewidth}|}
\hline
D                                                                                       & Detection Options                                                                                                                                                                                                                                                                                                     & Adaptation Strategies                                                                                                                                                                                                                                                                                                                                                                                    & Specific Cases                                                                                                                                                                             \\ \hline

\multicolumn{1}{|l|}{
\multirow{1}{*}{\rotatebox[origin=c]{90}{\begin{tabular}[c]{@{}c@{}}\textbf{D1 - Road accident}\\\textbf{D4 - Another} \\\textbf{Road blockage} \end{tabular}\hspace{-1.4cm}}}} & \begin{tabular}[c]{@{}p{1\linewidth}@{}}• User feedback apps (Waze)\\     • C-ITS data from vehicles (V2I)\\     • Traffic information centers (API, C-ITS DENM, RDS TMC, …)\\     • Integrated rescue system dispatching\\     • Video detection (AI image processing)\\     • Traffic flow (TF) data from sensors\end{tabular} & \begin{tabular}[c]{@{}p{1\linewidth}@{}}• Warn road users nearby and with routes via the disturbance location\\     • Info + guidance at affected stops to MaaS alternatives\\     • Busses temporarily adjust routes if taxi/CAV can serve bypassed stops, if favorable \\     • Intelligent traffic signs, VMS and other devices adjust to changes in traffic – warnings, guidance, change signal plan\end{tabular} & \begin{tabular}[c]{@{}p{1\linewidth}@{}}• Other networks affected\\     • Reserved road lanes are affected\\     • Reaction varies based on road blockage\\     • Treat D4 as D2 when extended\end{tabular} \\ \hline
\multicolumn{1}{|l|}{
\multirow{1}{*}{\rotatebox[origin=c]{90}{\begin{tabular}[c]{@{}c@{}}\textbf{D2 –} \\\textbf{Planned} \\\textbf{Work zone} \end{tabular}\hspace{-0.7cm}}}}                                                                  & \begin{tabular}[c]{@{}p{1\linewidth}@{}}• Digitized registry of planned work zones with temporary traffic management measures\\     • Intelligent ``first/last cone" or contractor updates*\end{tabular}                                                                                                                   & \begin{tabular}[c]{@{}p{1\linewidth}@{}}• Temporarily adjust networks for all affected modes and react aptly             \end{tabular}                                                                                                                                                                                                                                                                                                                          & \begin{tabular}[c]{@{}p{1\linewidth}@{}}• Adjust duration through regular detection means if not updated dynamically \end{tabular}                                                                                                          \\ \hline
\multicolumn{1}{|l|}{
\multirow{1}{*}{\rotatebox[origin=c]{90}{\begin{tabular}[c]{@{}c@{}}\textbf{D3 –} \\\textbf{Unplanned} \\\textbf{Work zone} \end{tabular}\hspace{-0.7cm}}}}                                                                & \begin{tabular}[c]{@{}p{1\linewidth}@{}}• Integrated rescue system**\\     • User feedback apps (Waze)\\     • C-ITS data from vehicles \\     • Intelligent ``cone"*\\     • Traffic information centers\\     • TF data from sensors\end{tabular}                                 & \begin{tabular}[c]{@{}p{1\linewidth}@{}}• Mostly the same as D1\\     • For long-lasting work zones, treat as D2 once details are available\end{tabular}                                                                                                                                                                                                                                                              & \begin{tabular}[c]{@{}p{1\linewidth}@{}}• Mostly the same as D1\\     • Response varies depending on expected duration \end{tabular}                                                          \\ \hline
\multicolumn{1}{|l|}{
\multirow{1}{*}{\rotatebox[origin=c]{90}{\begin{tabular}[c]{@{}c@{}}\textbf{D5-D7 –} \\\textbf{Broken vehicle} \\\textbf{Rail block}\end{tabular}\hspace{-1cm}}}}                                                       & \begin{tabular}[c]{@{}p{1\linewidth}@{}}• C-ITS data from vehicles (V2I)\\     • Railway or public transport (PT) operator dispatching\\     • Otherwise treat as D1/D4\end{tabular}                                                                                                                                                    & \begin{tabular}[c]{@{}p{1\linewidth}@{}}• Inform passengers about alternatives within the MaaS\\     • Inform M-ITS about the obstacle and react accordingly (mix of D1 and D3)\\     • For D6, D7, replace operation with busses/CAVs\end{tabular}                                                                                                                                                                   & \begin{tabular}[c]{@{}p{1\linewidth}@{}}• Certain PT routes (e.g. airport line) have higher priority \\     • Follow train/ metro safety standards\end{tabular}                                                         \\ \hline
\multicolumn{1}{|l|}{
\multirow{1}{*}{\rotatebox[origin=c]{90}{\begin{tabular}[c]{@{}c@{}}\textbf{D8 -} \\\textbf{Broken}\\\textbf{Traffic signals} \end{tabular}\hspace{-0.9cm}}}}                                                                 & \begin{tabular}[c]{@{}p{1\linewidth}@{}}• C-ITS (I2I) from the device\\     • Device operator\\     • TF data from sensors\end{tabular}                                                                                                                                                                                            & \begin{tabular}[c]{@{}p{1\linewidth}@{}}• Inform police to regulate traffic at the intersection \\     • Warn relevant road users\\     • Optimize nearby traffic signals for changed traffic\\     • Adjust routing for MaaS\end{tabular}                                                                                                                                                                                                     & \begin{tabular}[c]{@{}p{1\linewidth}@{}}• Warn trams, if affected, and react suitably\end{tabular}                                                                                                                                                \\ \hline
\multicolumn{1}{|l|}{
\multirow{1}{*}{\rotatebox[origin=c]{90}{\begin{tabular}[c]{@{}c@{}}\textbf{D9 -} \\\textbf{Rescue} \\\textbf{event}\end{tabular}\hspace{-0.4cm}}}}                                                                        & \begin{tabular}[c]{@{}p{1\linewidth}@{}}• C-ITS data from vehicles\\     • Integrated rescue system dispatching\end{tabular}                                                                                                                                                                                                 & \begin{tabular}[c]{@{}p{1\linewidth}@{}}• Warn nearby road users\\     • Re-route vehicles or adjust signal plans for smooth rescue\end{tabular}                                                                                                                                                                                                                                                                             & \begin{tabular}[c]{@{}p{1\linewidth}@{}}• Consider operation size and duration \end{tabular}                                                                                                                                                      \\ \hline
\multicolumn{4}{|l|}{\begin{tabular}[c]{@{}p{1\linewidth}@{}}*C-ITS-equipped ``traffic cone" or mobile app to update WZ start/end time.\\***Police should know first, communication with M-ITS may vary among countries.\end{tabular}}                                                                                                                                                                                                                                                                                                                                                                                                                                                                                                                                                                                                                                                                                                                                                                                                                                                                                                                                                                                                                                                          \\ \hline
\end{tabular}
\end{table}

Additional causes of disturbance include major sport or cultural events, and gatherings, that significantly increase (or decrease) traffic demand within certain areas of the networks. Therefore, a registry of such events, including information about the event, time, location, affected area, and possibly expected number of visitors should be included.
Some disturbances, especially the external ones mentioned above, can be observed and identified centrally. Then only the warnings with relevant information can be shared with relevant individual edge devices to be evaluated individually. This includes the processing of information about detected disturbances from the edge devices and transport users in cases where the impacts of the disturbance exceed the communication range of the local edge devices.

\textbf{Other impact mitigation suggestions are as follows:}
--For damaged CAVs, inform its users about nearby replacements and send warnings if blocking the road.
--For broken shared bikes and scooters, notify users of alternatives; other modes are not affected. --Temporarily adjust service limits for space-limited services when intermodal nodes near the borders are out of service.
--Warn users at affected PT stops and guide them to nearby alternatives or other MaaS services.
--Empty bike-sharing nests provide info about nearby nests with available bikes and PT alternatives.
--Notify taxis and CAVs of disturbances and areas with increased demand.
--Provide (multimodal) route change suggestions for users with planned routes.
--General control measures for improved performance: Integrate (multimodal) navigation apps into M-ITS; Increase C-ITS systems and OBU penetration; Establish disturbance detection information systems; Implement VMS and variable LED signage; Use intelligent traffic lights, including cooperation, PT prioritization, and multimodal; Create incentives (cost evaluation, tolling, gamification); Implement multimodal ticket schemes; Consider multimodal performance in traffic control systems; Design new data standards for interoperability across the EU (identify best practices for data management; define minimal operational standards for data requirements).
\vspace{-10pt}
\section{Data Sharing at the Network Edge}
\vspace{-10pt}

The disturbance-mode-network model defines combinations of disturbances and affected transportation modes/networks, but not all users or vehicles of the affected mode-network combination must be warned. Only relevant edge devices should be informed with specific details. Given the small size of the basic warning data package, this can be spread quite broadly with more details given upon request, if available. Disturbance warnings sent to edge devices should include the following information, --Type of disturbance, --Location on the network affected modes, and class of the network segment for each mode (e.g., critical, major, inferior, minor), --Expected impact severity (one or more measures as applicable): Capacity reduction; Lanes affected; Severity index; Displaced traffic volume, --Estimated time of duration, and any other case-specific information.

To efficiently distribute warnings for disturbances, it is crucial to consider various parameters like the affected area, specific modes, networks, and user classes. The goal is to notify only relevant users and devices without overwhelming them. This involves a combination of general heuristic rules for area-based warnings, C-ITS traces for approaching vehicles, data from edge devices regarding their planned routes (including non-scheduled traffic), adaptation strategies for dealing with the disturbance, and machine learning algorithms for predicting trajectories. There are two main aspects to consider: (i) information about the disturbance, which is part of the warning itself (type, location, severity, duration), and (2) information about the recipient of the warning. For recipients, it is essential to account for their expected or planned trajectory, any expected TF changes along that trajectory, and whether they are part of the adaptation strategy.

Edge devices in this environment encompass a wide range of components, from sensors and processors within vehicles, through users' mobile phones, to infrastructure elements like traffic signal controllers and communication nodes. Edge computing can greatly enhance the efficiency of distributing warnings for disturbances based on the following considerations: \textbf{Efficient Data Processing:} Edge computing processes data closer to the source, enabling quick assessment of disturbances and reducing response times. For example, a CAV detects a fallen tree on the road and immediately sends an alert to nearby vehicles.
\textbf{Localized Decision-Making:} Edge devices can make localized decisions based on the available information. They can analyze their planned routes, expected TF changes, and other relevant factors to determine whether a particular disturbance warning is relevant to them and address it accordingly.
\textbf{Adaptation Strategies:} Edge devices can autonomously implement the pre-defined adaptation strategies and cooperate with other devices to optimize overall performance.
\textbf{Reduced Communication Network Congestion:} By processing and filtering data at the edge, the system can significantly reduce congestion of the communication network. This is crucial during disturbances when there may be an increased volume of data traffic due to widespread alerts.     

\vspace{-10pt}
\section{Conclusions}
This paper has explored the critical role of M-ITS in enhancing MaaS efficiency, especially during disturbances. Establishing relationships between transportation modes, networks, and disturbances, has highlighted the importance of data sharing on the network edge. The selective distribution of disturbance warnings to relevant users and devices is crucial for effective disturbance mitigation. This framework offers valuable insights for building resilient and adaptive urban transportation systems that prioritize sustainability and user needs.
\vspace{-10pt}
\section{Acknowledgements}
This project has received funding from the RE-ROUTE Project, the European Union's Horizon Europe research and innovation program under the Marie Skłodowska-Curie grant agreement No 101086343.
\vspace{-10pt}
\bibliographystyle{styles/bibtex/splncs_srt.bst}

\end{document}